\documentstyle[12pt,epsfig,axodraw,a4]{article} 
\def\bild#1#2{    
        \vspace*{-5mm}
        \begin{center}
        \begin{math}
        \epsfxsize#2cm
        \epsffile{#1}
        \end{math}
        \end{center}  }
\newcommand{\vs}{\vspace{-0.25cm}}
\begin{document} 
\begin{center}
\large{\bf Parity-violating two-pion exchange nucleon-nucleon interaction}

\medskip

N. Kaiser\\

\smallskip

{\small Physik Department T39, Technische Universit\"{a}t M\"{u}nchen,
    D-85747 Garching, Germany}

\end{center}

\bigskip

\begin{abstract}
We calculate in chiral perturbation theory the parity-violating two-pion
exchange nucleon-nucleon potentials at leading one-loop order. At a distance 
of $r= m_\pi^{-1} \simeq 1.4\,$fm they amount to about $\pm 16\%$ of the 
parity-violating $1\pi$-exchange potential. We evaluate also the 
parity-violating effects arising from $2\pi$-exchange with excitation of 
virtual $\Delta(1232)$-isobars. These come out to be relatively small in 
comparison to those from diagrams with only nucleon intermediate states. The 
reason for this opposite behavior to the parity-conserving case is the 
blocking of the dominant isoscalar central channel by CP-invariance. 
Furthermore, we calculate the T-matrix related to the iteration of 
the parity-violating $1\pi$-exchange with the parity-conserving one. The
analytical results presented in this work can be easily implemented into 
calculations of parity-violating nuclear observables.      

\end{abstract}


\bigskip


\bigskip

Nuclear parity violation is an important tool to study the standard model 
of strong and electroweak interactions \cite{adel,musolf}. In the two-nucleon 
system parity violation has been traditionally represented by parity-violating 
one-meson exchange where the strong and weak interactions are parametrized 
through parity-conserving and parity-violating meson-nucleon vertices. Due to
CP-invariance, there exists no parity-violating coupling of neutral scalar or
pseudoscalar mesons to nucleons (Barton's theorem). Therefore, standard 
parametrizations of the parity-violating NN-potential involve
the exchange of charged pions ($\pi^\pm$) as well as vector mesons
($\rho^{\pm,0}$ and $\omega$). The pertinent parity-violating meson-nucleon
coupling constants $h^{(0,1,2)}_{\pi,\rho,\omega} \sim 10^{-7}$ have been
calculated in quark models \cite{ddh} and soliton models \cite{soliton}. Due 
to a variety of uncertainties (e.g. quark wavefunctions, strong interaction 
enhancements, choice of particular soliton model) only ``reasonable'' ranges 
could be derived so far. For the parity-violating $\pi NN$-coupling constant 
$h_\pi$ an experimental upper bound is known to be $|h_\pi| < 1.43 \cdot 
10^{-7}$ \cite{adel}. An improved determination of this coupling constant is 
expected  from a measurement of the photon asymmetry $A_\gamma$ in the 
radiative capture of thermal neutrons $\vec n p \to d \gamma$ 
\cite{hyun,kaplan,npdg}.

Recently, nuclear parity violation has been reformulated in the framework of
effective field theory \cite{pveft,liu}. As in the case of the 
parity-conserving NN-interaction one exploits the separation of
scales. At low energies only the pions are kept as explicit degrees of freedom
while all heavier particles are integrated out and their dynamical effects are
subsumed in contact terms. At very low energies $E_{cm} < 10\,$MeV even the
pion can be integrated out and then one is working with the pionless version
of the effective field theory for nuclear parity violation. It has been shown
in ref.\cite{pveft} that at leading order ${\cal O}(q)$ there are in total
five parity-violating low-energy constants associated with the respective
contact operators linear in the nucleon momenta. One therefore requires a
minimum of five independent, low-energy observables. For the effective field
theory with dynamical pions there appear up to order ${\cal O}(q)$ three more
parameters: the parity-violating $\pi N$-coupling constant $h_\pi$, a 
next-to-next-to-leading order correction to it, and a new electromagnetic 
operator \cite{pveft}. This new electromagnetic operator is a specific feature
of the systematic effective field theory framework and entirely absent in the 
one-meson exchange phenomenology.  

Clearly, a lot of work is still necessary e.g. in order to pin down the 
parity-violating low-energy constants and thus to reach a level where the 
effective field theory framework becomes predictive. In that situation it
seems worthwhile to investigate separately the hierarchy of long-range
pion-induced  parity-violating NN forces. This is the  purpose of 
the present short paper. We will compare directly the parity-violating
$2\pi$-exchange potentials at leading one-loop order with the parity-violating
$1\pi$-exchange. For the parity-conserving NN-potential the $2\pi$-exchange 
with excitation of the low-lying spin-isospin-$3/2$  $\Delta(1232)$-resonance
plays a major role. The corresponding potentials exceed the ones from
diagrams with only nucleon intermediate states typically by an order of
magnitude \cite{nnpot,krebs}.  (Recently, is has been found that this feature
remains when electromagnetic (one-photon exchange) corrections are included
\cite{pipigamma}.) Therefore it is important to check whether a similar
dominance of the $\Delta$-induced processes holds for the parity-violating 
$2\pi$-exchange interaction. We find that this is not the case. The basic 
reason for this opposite behavior is CP-invariance which forbids 
parity-violating effect in the isoscalar central channel. As a further
possibly sizeable one-loop contribution we calculate the T-matrix related to
the iteration of the parity-violating $1\pi$-exchange with the
parity-conserving one. 

Let us begin with recalling the parity-violating pion-nucleon vertex. It has
the form:
\begin{equation} {\cal L}_{\rm pv} = {h_\pi \over \sqrt{2}} \, \bar N
 (\vec \pi \times \vec \tau\,)^3 N \,,\end{equation}
where $N$ denotes a nucleon Dirac-spinor and $h_\pi\sim 10^{-7}$ is the weak
$\pi NN$-coupling constant. In the center-of-mass frame the T-matrix of 
parity-violating $1\pi$-exchange is readily computed as:   
\begin{equation} {\cal T}^{(1\pi)}_{\rm pv}=-i \,{g_A h_{\pi}\over 2 
\sqrt{2} f_\pi} (\vec \tau_1 \times \vec \tau_2)^3 \,\, { (\vec \sigma_1+\vec
\sigma_2) \cdot \vec q \over m_\pi^2 +q^2}+{\cal O}(M_N^{-2})\,. \end{equation}
Here, $g_A \simeq 1.3$ is the nucleon axial vector coupling constant and 
$f_\pi = 92.4 \,$MeV denotes the pion decay constant. $\vec \sigma_{1,2}$ and
$\vec \tau_{1,2}$ are the usual spin- and isospin-operators of the two 
nucleons and $\vec q$ stands for the momentum transfer. Note that relativistic 
corrections start at order $M_N^{-2}$ (with $M_N=939\,$MeV the nucleon mass) 
and therefore are negligibly small (typically less than $1\%$). Fourier 
transformation of the static term in Eq.(2) leads to a parity-violating 
NN potential in coordinate  space: 
\begin{equation} V_{\rm pv}(\vec r\,) = (\vec \tau_1 \times \vec \tau_2)^3 \,
(\vec \sigma_1+\vec \sigma_2)\cdot \hat r \,\, U(r) +  (\tau_1^3 + \tau_2^3) 
(\vec \sigma_1\times \vec \sigma_2)\cdot \hat r \,\, W(r) \,, \end{equation}
with the radial dependence given by the derivative of a Yukawa function:
\begin{equation} U(r)^{(1\pi)} = - {g_A h_{\pi}\over 8 \sqrt{2} \pi f_\pi} \,
  {e^{-m_\pi r} \over r^2} (1+m_\pi r) \,. \end{equation}
The second term in Eq.(3) proportional to the cross product $\vec \sigma_1
\times \vec \sigma_2$ of spin-operators has also been introduced because the 
associated ``vector-type'' potential $W(r)$ does receive contributions 
from parity-violating $2\pi$-exchange.   

Next, we come to the parity-violating two-pion exchange. Some representative 
diagrams with leading order vertices are shown in Fig.\,1. These are to be 
supplemented by further diagrams with the parity-violating vertex at a 
different position and/or both nucleon lines interchanged. We are interested
only in the nonpolynomial or finite-range parts of these one-loop diagrams
(disregarding the zero-range $\delta^3(\vec r\,)$-terms from the polynomial
pieces). For that purpose it is sufficient to calculate their spectral 
functions or imaginary parts using the Cutkosky cutting rule. The pertinent
two-body phase space integral is most conveniently performed in the $\pi\pi$
center-of-mass frame where it becomes proportional to a simple angular
integral: $\sqrt{\mu^2 -4m_\pi^2}/(32\pi \mu) \int_{-1}^1 dx$, with $\mu >
2m_\pi$ the $\pi\pi$ invariant mass. Using this technique we get from the
leading one-loop diagrams  in Fig.\,1:

\bigskip
\bild{semfig2.epsi}{12}
\vskip -0.3cm
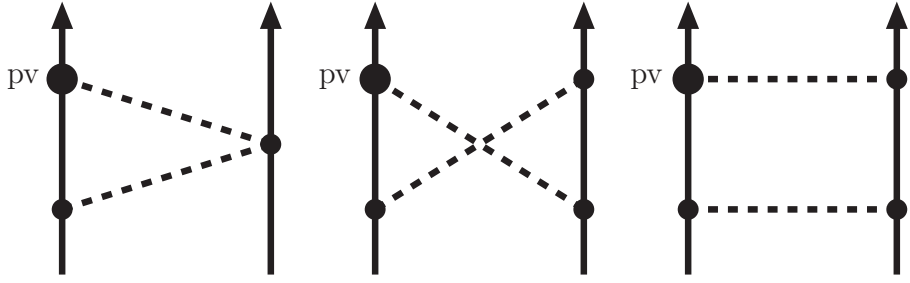
{\it Fig.\,1: Diagrams related to parity-violating two-pion exchange. The 
heavy dot symbolizes the parity-violating pion-nucleon vertex: $i\, h_\pi 
\epsilon^{ab3} \tau^b/\sqrt{2}$. Diagrams with the parity-violating vertex at 
a different position are not shown.} 
\bigskip

\begin{eqnarray} {\rm Im} {\cal T}^{(2\pi)}_{\rm pv} &=&{g_A h_{\pi}\sqrt{
\mu^2-4m_\pi^2} \over 64 \sqrt{2}\pi f_\pi^3\, \mu}\Bigg\{ i(\vec \tau_1\times 
\vec \tau_2)^3 \,(\vec \sigma_1+\vec \sigma_2)\cdot \vec q \, \bigg[ g_A^2{3 
\mu^2-8 m_\pi^2 \over \mu^2- 4m_\pi^2} -1 \bigg]   \nonumber \\ && \qquad\qquad
\qquad \qquad -4 g_A^2\, i (\tau_1^3+ \tau_2^3) \, (\vec \sigma_1\times \vec 
\sigma_2) \cdot \vec q \, \Bigg\} \,. \end{eqnarray}
The notation Im\,$T^{2\pi}_{\rm pv}$ is meant here such that one is taking the
imaginary part of the loop functions standing to the right of the spin- and 
isospin factors. For the box diagrams with two nucleon propagators one
encounters the nontrivial angular integral: $\int_{-1}^1 dx\,(x-i 0)^{-2}=-2$. 
The one-loop T-matrix in momentum space can be reconstructed from the spectral 
function via a once-subtracted dispersion relation:
\begin{equation} {\cal T}^{(2\pi)}_{\rm pv}  = -{2q^2 \over \pi} \int_{2m_\pi
}^\infty d\mu \,{{\rm Im} {\cal T}^{(2\pi)}_{\rm pv}\over \mu(\mu^2 + q^2)}\,,
\end{equation} 
and it agrees (up to an irrelevant subtraction constant) with the result of
ref.\cite{pveft} (see Eqs.(120-122) therein). With the help of the mass
spectra in Eq.(5) one can also calculate directly the parity-violating 
$2\pi$-exchange potential in coordinate space. One finds the following radial 
dependences of the scalar-type potential:  
\begin{equation} U(r)^{(2\pi)}={g_A h_{\pi} m_\pi\over \sqrt{2}(4\pi f_\pi 
r)^3} \,\Big\{ (8 g_A^2-2)m_\pi r \, K_0(2 m_\pi r) +(4 g_A^2m_\pi^2 r^2+9
g_A^2-3) \, K_1(2 m_\pi r)  \Big\}  \,, \end{equation}
and the vector-type potential:
\begin{equation}W(r)^{(2\pi)}=-{g_A^3h_{\pi} m_\pi\over(2\sqrt{2}\pi f_\pi 
r)^3} \,\Big\{ 2m_\pi r \,K_0(2 m_\pi r)+3\, K_1(2 m_\pi r)\Big\} \,,
\end{equation} 
with $K_{0,1}(2 m_\pi r)$ two modified Bessel functions. Their asymptotic
behavior for large distances $r$ is: $U(r)^{(2\pi)}\sim e^{-2m_\pi r}r^{-3/2}$
and $W(r)^{(2\pi)}\sim e^{-2m_\pi r} r^{-5/2}$. Moreover, in the chiral limit 
$m_\pi=0$  these potentials follow a simple $r^{-4}$ form, a feature which
could also be guessed by counting mass dimensions.

\vskip -0.3cm
\begin{table}[hbt]
\begin{center}
\begin{tabular}{|c|cccccccccc|}
\hline 
$r$~[fm]& 1.0 & 1.1 & 1.2 & 1.3 & 1.4 & 1.5 & 1.6 & 1.7 & 1.8 & 1.9 \\ 
\hline
$U^{(2\pi)}/U^{(1\pi)}$ & --0.332 &  --0.271 & --0.224 & --0.188 &
--0.159 & --0.136 & --0.117 & --0.102 & --0.089 & --0.078 \\
$W^{(2\pi)}/U^{(1\pi)}$ & 0.420 & 0.331 & 0.265 & 0.215 & 0.177  &
0.146 & 0.122 &  0.103 & 0.087 & 0.074   \\ \hline
$U^{(\Delta)}/U^{(1\pi)}$ & --0.190 & --0.134 & --0.097 & --0.072 &
--0.055  & --0.042 & --0.033 &  --0.026 & --0.021 & --0.017   \\
$W^{(\Delta)}/U^{(1\pi)}$ & 0.322 & 0.229 & 0.167 & 0.125  & 0.095 & 
0.073 & 0.057 & 0.045 & 0.036 & 0.029  \\ 
\hline
\end{tabular}
\end{center}
\vskip -0.2cm
{\it Table I: Ratio of parity-violating  $2\pi$-exchange potentials to the 
parity-violating $1\pi$-exchange potential $U(r)^{(1\pi)}\sim e^{-m_\pi r}(1+ 
m_\pi r)/r^2$ as a function of  the  nucleon distance $r$.}
\end{table}

We are now in the position to present numerical results. The numbers in the 
first and second row of Table\,I  give the ratio between the parity-violating
$2\pi$-exchange potentials $U(r)^{(2\pi)}$ and  $W(r)^{(2\pi)}$ and the 
parity-violating $1\pi$-exchange potential $U(r)^{(1\pi)}$ for distances 
$1.0\,{\rm fm}\leq r\leq1.9\,{\rm fm}$ (see also Fig.\,2). One sees that in
the region around the pion Compton wavelength  $r =m_\pi^{-1} \simeq 1.4\,{\rm
fm}$ this ratio varies between $10\%$ and $25\%$ (with opposite sign for the
scalar and vector type potential). Such a suppression of the parity-violating
$2\pi$-exchange interaction as deduced here from the ratio of coordinate space
potentials is quantitatively consistent with the results of ref.\cite{hyun}. 
There the combined effect of a cut-off regularized $2\pi$-exchange,
short-distance counterterms and the new electromagnetic operator on the photon
asymmetry $A_\gamma(\vec n p \to d \gamma)$ has been quantified as a (minus)
$10 \sim 20 \%$ correction to the dominant one-pion exchange contribution.   
         
\vskip 1.0cm

\bigskip
\bild{pvpot.eps}{10}
\vskip 0.2cm
{\it Fig.\,2: Ratios of parity-violating $2\pi$-exchange potentials to the 
parity-violating $1\pi$-exchange.}

\vskip 0.2cm

\bigskip
\bild{semfig3.epsi}{8}
\vskip -0.3cm
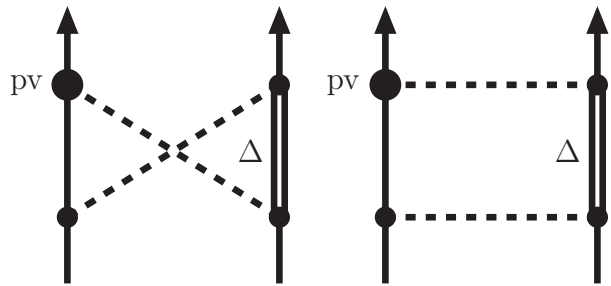
{\it Fig.\,3: Diagrams related to parity-violating $2\pi$-exchange with
single $\Delta(1232)$-isobar excitation.}
\bigskip

As mentioned in the introduction the parity-conserving $2\pi$-exchange
interaction does not come primarily from leading order diagrams as in Fig.\,1. 
The by far dominant contribution arises from processes where the low-lying 
$\Delta(1232)$-resonance is excited in the intermediate state \cite{nnpot}. 
Equivalently, one can obtain this dominant contribution from $\pi\pi NN$
contact vertices proportional to the large low-energy constants $c_3$ and
$c_4$ (which effectively include the important $\Delta$-dynamics). It is
therefore important to check whether a similar dominance of the
$\Delta$-induced processes holds for the parity-violating $2\pi$-exchange 
interaction. Fig.\,3 shows the relevant $2\pi$-exchange diagrams with single
$\Delta(1232)$-isobar excitation. We do not consider a parity-violating $\pi N
\Delta$-coupling since it is of higher order than the  parity-violating $\pi
NN$-vertex: $i\, h_\pi \epsilon^{ab3} \tau^b/\sqrt{2}$ (the vectorial $(1/2
\leftrightarrow 3/ 2)$ spin-transition operator must be contracted 
with a momentum in order to ensure rotational invariance). Using the
abovementioned technique to calculate the spectral function we find from the
$2\pi$-exchange diagrams in Fig.\,3:   
\begin{eqnarray} {\rm Im} {\cal T}^{(\Delta)}_{\rm pv} &=& {g_A^3 h_{\pi} 
\over 64 \sqrt{2} \pi f_\pi^3\,\mu}\Bigg\{ i(\vec\tau_1\times \vec \tau_2)^3 \,
(\vec \sigma_1+\vec \sigma_2) \cdot \vec q \, \Bigg[ -\sqrt{\mu^2-4m_\pi^2}
 + {1 \over \Delta} (\mu^2 +2\Delta^2-2m_\pi^2) \nonumber \\ && \times\arctan{
\sqrt{\mu^2-4m_\pi^2}\over 2\Delta} \Bigg]  + i (\tau_1^3+ 
\tau_2^3) \, (\vec \sigma_1\times \vec \sigma_2)\cdot \vec q \, \Bigg[{ \pi 
\over 4\Delta}(4m_\pi^2 - \mu^2) \nonumber \\ && +2\sqrt{ \mu^2-4m_\pi^2} 
+ {1 \over \Delta}(4m_\pi^2-4 \Delta^2-\mu^2) \arctan{\sqrt{\mu^2-  
4m_\pi^2}\over 2 \Delta}\Bigg] \Bigg\} \,, \end{eqnarray} 
where the term $\arctan(\sqrt{\mu^2- 4m_\pi^2}/2 \Delta)$ is typical for a
$\Delta(1232)$-isobar propagating in a (pion) loop \cite{nnpot}. Here, 
$\Delta =293\,$MeV denotes the delta-nucleon mass splitting. It is treated as 
a small scale comparable to the pion mass $m_\pi$ and the momentum transfer
$\vec q$. In Eq.(9) we have already inserted the empirically well-satisfied
relation $g_{\pi N \Delta} = 3 g_A M_N/\sqrt{2}f_\pi$ for the $\pi N
\Delta$-coupling constant. With the help of the spectral Im${\cal T}^{(\Delta)
}_{\rm pv}$ one can calculate the T-matrix in momentum space, using Eq.(6),
or the scalar and vector type potentials in coordinate space:
\begin{eqnarray} U(r)^{(\Delta)}&=& {g_A^3 h_{\pi}\over(4\sqrt{2}\pi f_\pi)^3
r^2} \int_{2 m_\pi}^\infty d\mu \, e^{-\mu r} (1+\mu r) \Bigg[-\sqrt{\mu^2-4 
m_\pi^2} \nonumber \\ && +{1 \over \Delta} (\mu^2 +2\Delta^2-2m_\pi^2)
\arctan{ \sqrt{\mu^2-4m_\pi^2}\over 2\Delta} \Bigg] \,, \end{eqnarray}
\begin{eqnarray} W(r)^{(\Delta)}&=& {g_A^3 h_{\pi}\over(4\sqrt{2}\pi f_\pi)^3
r^2} \int_{2 m_\pi}^\infty d\mu \, e^{-\mu r} (1+\mu r)\Bigg[{\pi\over4\Delta}
(4m_\pi^2 -\mu^2)  \nonumber \\ && + 2\sqrt{\mu^2- 4m_\pi^2} + {1\over \Delta}
(4m_\pi^2-4 \Delta^2-\mu^2) \arctan{\sqrt{\mu^2-4m_\pi^2}\over 2\Delta}\Bigg] 
\, \end{eqnarray} 
The first term in square brackets of E.(11) is the leading one in an expansion 
in powers of $1/\Delta$ and it makes up about $60\%$ of the total vector type
potential $W(r)^{(\Delta)}$. The functional form of this piece reads: $-e^{-2
m_\pi r}(1+m_\pi r)^2/r^5$. It is also interesting to note that if one would 
work with $\pi\pi NN$ contact vertices proportional to the large low-energy
constants $c_{3,4}$ (which represent partly the $\Delta$-dynamics) one would
get a nonvanishing contribution only from $c_4$. It corresponds precisely to 
this leading $1/\Delta$ term just mentioned.  

Let us again turn to numerical results. The numbers in the third and fourth 
row of Table\,I  give the ratio between the parity-violating $2\pi$-exchange 
potentials $U(r)^{(\Delta)}$ and $W(r)^{(\Delta)}$ and the parity-violating 
$1\pi$-exchange potential $U(r)^{(1\pi)}$. First, one observes that the 
$\Delta$-induced potentials are of the same sign as the ones from diagrams
with only nucleon intermediate states. Secondly, they are suppressed by a 
factor 2 to 4 (in the region around the pion Compton wavelength $r=m_\pi^{-1}
$. This behavior is completely opposite to the parity-conserving case. The 
primary reason for that is CP-invariance which excludes parity-violating 
effects in the isoscalar central channel (Barton's theorem). We note once more 
that the $c_3$ contact vertex which is solely responsible for strong isoscalar 
central NN-attraction leads to a {\it vanishing} contribution to the
parity-violating $2\pi$-exchange. We can therefore conclude that the 
calculation of ref.\cite{hyun} which aims at an improved determination of the 
weak $\pi NN$-coupling constant $h_\pi$ from the photon asymmetry $A_\gamma(
\vec n p \to d \gamma)$ is not affected much by the inclusion of
$\Delta$-induced $2\pi$-exchange corrections. 

The (right) planar box diagram in Fig.\,1 contains also a two-nucleon
reducible piece: the iteration of the parity-violating $1\pi$-exchange with
the parity-conserving $1\pi$-exchange. It is distinguished by the feature
that its energy denominator is given by the difference of nucleon kinetic
energies. This enhances the corresponding T-matrix by a large scale factor,
the nucleon mass $M_N=939\,$MeV. For the center-of-mass kinematics $N_1(\vec 
p\,)+N_2(-\vec p\,) \to N_1(\vec p\,')+N_2(-\vec p\,')$ we obtain the 
following expression for the T-matrix:    
\begin{eqnarray} {\cal T}_{\rm pv} &=& {g_A^3 h_\pi M_N \over \pi \sqrt{2}(4
f_\pi)^3} \bigg\{ (\vec \tau_1\times \vec \tau_2)^3 \bigg[ i (\vec \sigma_1+
\vec \sigma_2)\cdot \vec q \,\Big((2m_\pi^2+q^2)\, G_0 -2 \Gamma_0\Big)
\nonumber \\ && +  (\vec \sigma_1\cdot \vec q \,\,\vec \sigma_2 \cdot \vec q 
\times \vec p +\vec\sigma_2\cdot \vec q \,\,\vec \sigma_1 \cdot \vec q \times 
\vec p\,)\Big(2G_0+4G_1\Big) \bigg] \nonumber \\ && +2(\tau_1^3-\tau_2^3) 
\bigg[ (\vec \sigma_1+ \vec \sigma_2)\cdot (\vec p + \vec p\,') \Big(2\Gamma_0
+2\Gamma_1 -(2m_\pi^2+q^2)(G_0+2G_1)\Big) \nonumber \\ && +  i (\vec \sigma_1
\cdot (\vec p + \vec p\,')\, \vec \sigma_2 \cdot \vec q \times \vec p + \vec 
\sigma_2\cdot (\vec p + \vec p\,')\, \vec \sigma_1 \cdot  \vec q \times \vec p 
\,) \Big(2G_0+8G_1+8G_3\Big) \bigg] \bigg\} \,, \end{eqnarray} 
where $p = |\vec p\,|= |\vec p\,'|$ is the center-of-mass momentum and $\vec q 
=\vec p\,'-\vec p$ the momentum transfer. The occurring complex-valued loop 
functions read:
\begin{equation} \Gamma_0(p) = {1\over 4p } \bigg[ 2\arctan{2p \over m_\pi} +
  i \, \ln\bigg( 1 + {4p^2 \over m_\pi^2} \bigg) \bigg] \,,\end{equation} 
\begin{equation} \Gamma_1(p)={1\over 2p^2} \bigg[ m_\pi + i \, p
  -(m_\pi^2+2p^2) \Gamma_0(p) \bigg] \,,\end{equation}
\begin{eqnarray} G_0(p,q) &=& {1\over q \sqrt{m_\pi^4 +p^2(4m_\pi^2+q^2)}} 
\Bigg[\arcsin{  q \,m_\pi \over \sqrt{(m^2_\pi +4p^2)(4m_\pi^2 +q^2)}} \\ 
\nonumber &&+ \, i \, \ln{ p\, q + \sqrt{m_\pi^4 +p^2(4m_\pi^2+q^2)} \over 
m_\pi \sqrt{m_\pi^2 +4p^2}}\Bigg] \,, \end{eqnarray} 
\begin{equation}  G_1(p,q) = { \Gamma_0(p)- q^{-1} \arctan{q \over 2m_\pi}
    -(m_\pi^2+2p^2)  G_0(p,q) \over 4p^2-q^2} \,, \end{equation}
\begin{equation}  G_3(p,q) = { {1\over 2}\Gamma_1(p)-p^2  G_0(p,q)-2(m_\pi^2+
2p^2)  G_1(p,q) \over 4p^2-q^2}  \,,\end{equation}
where $\Gamma_{0,1}(p)$ originate from loop integrals with one pion propagator
and $G_{0,1,3}(p,q)$ from loop integrals with two pion propagators. 
Interestingly, the one-loop T-matrix ${\cal T}_{\rm pv}$ in Eq.(12) related to 
iterated $1\pi$-exchange is ultraviolet convergent, since the parity-violating 
$\pi N$-vertex is momentum independent.  ${\cal T}_{\rm pv}$ represents a 
complex-valued unitarity correction which cannot be Fourier-transformed into a 
coordinate space potential. This inhibits a direct comparison with the
parity-violating $1\pi$- and $2\pi$-exchange potentials. It should also be 
noted that ${\cal T}_{\rm pv}$ is partly accounted for by the use of realistic 
nucleon wavefunctions in perturbative calculations of parity-violating
observables. Nevertheless, its explicit inclusion may be desirable under
certain circumstances.           
            


In summary, we have calculated in this work the long-range tail of the 
parity-violating $2\pi$-exchange interaction. The leading order diagrams with
only nucleon intermediate states amount to a $10-20\%$ correction of the
parity-violating $1\pi$-exchange. The diagrams with excitation of virtual
$\Delta$-isobars are significantly suppressed. This completely opposite
behavior to the parity-conserving case is a consequence of CP-invariance which
forbids parity-violating effects in the isoscalar central channel. The
analytical results presented in this work can be easily implemented into
calculations of parity-violating nuclear observables.


\begin{thebibliography}{99}
\bibitem{adel} E.G. Adelberger and W.C. Haxton, {\it Ann. Rev. Nucl. Part. 
Sci.}  {\bf 35}, 501 (1985).\vs
\bibitem{musolf} M.J. Ramsey-Musolf and S.A. Page, {\it Ann. Rev. Nucl. Part. 
Sci.} {\bf 56}, 1 (2006); and refs. therein.\vs
\bibitem{ddh} B. Desplanques, J.F. Donoghue, and B.R. Holstein, {\it Ann. Phys.
(NY)} {\bf 124}, 449 (1980); and refs. therein.\vs
\bibitem{soliton} N. Kaiser and Ulf-G. Mei{\ss}ner, {\it Nucl. Phys.} {\bf
A499}, 699 (1989).\vs
\bibitem{hyun} C.H. Hyun, S. Ando and  B. Desplanques, {\it Eur. Phys. J.}
{\bf A32}, 513 (2007); {\it Phys. Lett.} {\bf B651}, 257 (2007).\vs
\bibitem{kaplan} D.B. Kaplan, M.J. Savage, R.P. Springer, and  M.B. Wise, {\it 
Phys. Lett.} {\bf B449}, 1 (1999).\vs
\bibitem{npdg} W.M Snow et al., {\it Nucl. Inst. Meth.} {\bf A440}, 729
(2000).\vs 
\bibitem{pveft} Shi-Lin Zhu, C.M. Maekawa, B.R. Holstein, M.J. Ramsey-Musolf 
and U. van Kolck, {\it Nucl. Phys.} {\bf A748}, 435 (2005); and refs.
therein.\vs
\bibitem{liu} C.P. Liu, {\it Phys. Rev.} {\bf C75}, 065501 (2007).\vs 
\bibitem{nnpot} N. Kaiser, S. Gerstend\"orfer, and W. Weise, {\it Nucl. Phys.} 
{\bf A637}, 395 (1998).\vs
\bibitem{krebs} H. Krebs, E. Epelbaum, and Ulf-G. Mei{\ss}ner, {\it Eur. Phys. 
J.} {\bf A32}, 127 (2007); and refs. therein.\vs
\bibitem{pipigamma} N. Kaiser, {\it Phys. Rev.} {\bf C73}, 064003 (2006); 
{\it Eur. Phys. J.} {\bf A31}, 207 (2007).\vs
\end{thebibliography}
\end{document}